*Classification:* Physical Sciences (Astronomy) and Biological Sciences (Biophysics)

# Detection of Circular Polarization in Light Scattered from Microbes

William B. Sparks[1]; James Hough[2]; Thomas A. Germer[3]; Feng Chen[4]; Shiladitya DasSarma[4]; Priya DasSarma[4]; Frank T. Robb[4]; Nadine Manset[5]; Ludmilla Kolokolova[6]; Neill Reid[1]; F. Duccio Macchetto[1]; William Martin[2]

[1] *Space Telescope Science Institute, 3700 San Martin Drive, Baltimore, MD 21218, USA*
[2] *Center for Astrophysics Research, University of Hertfordshire, UK*
[3] *National Institute of Standards and Technology, 100 Bureau Drive, Gaithersburg, MD 20899, USA*
[4] *Center of Marine Biotechnology, University of Maryland Biotechnology Institute, 701 East Pratt Street, Suite 236, Baltimore, MD 21202, USA*
[5] *Canada-France-Hawaii Telescope, 65-1238 Mamalahoa Highway, Kamuela, HI 96743, USA*
[6] *Department of Astronomy, University of Maryland, College Park, MD 20742, USA*

*Corresponding author:*
William B. Sparks,
Space Telescope Science Institute,
3700 San Martin Drive,
Baltimore, MD 21218, USA.
Phone: 410-338-4843
Fax: 410-338-4767
Email: sparks@stsci.edu

## Abstract

The identification of a universal biosignature that could be sensed remotely is critical to the prospects for success in the search for life elsewhere in the universe. A candidate universal biosignature is homochirality, which is likely to be a generic property of all biochemical life. Due to the optical activity of chiral molecules, it has been hypothesized that this unique characteristic may provide a suitable remote sensing probe using circular polarization spectroscopy. Here, we report the detection of circular polarization in light scattered by photosynthetic microbes. We show that the circular polarization appears to arise from circular dichroism of the strong electronic transitions of photosynthetic absorption bands. We conclude that circular polarization spectroscopy could provide a powerful remote sensing technique for generic life searches.

## *Introduction*

The search for life in the Universe depends on the identification of observable signatures that are unique to biological processes. If these signatures may be sensed remotely, then extensive surveys of planetary surfaces and distant objects may be undertaken without the need (initially) for costly landing spacecraft. A candidate universal biosignature that may lend itself to remote sensing application is homochirality, which, due to the optical activity of biological molecules, is potentially detectable using circular polarization spectroscopy (1-6). Organic molecules typically exist in two mirror-image forms and are said to be *chiral*; that is they exhibit handedness. All known living organisms use *only* left-handed or L-amino acids in proteins and right-handed or D-sugars in nucleic acids and this unique preference for just a single handedness is termed *homochirality*. Intriguingly, analysis of the Murchison meteorite has shown L-excesses of 2–9% for a number of α-methyl amino acids (7), with slightly smaller excesses found in the Murray meteorite (8). Homochirality is thought to be generic to all forms of biochemical life as a necessity for self-replication (9) and hence it is likely to be a signature of non-terrestrial life. To be detectable remotely using circular polarization, homochirality must imprint itself upon the circular polarization spectrum *in scattered light*. Here we report on the results of sensitive laboratory measurements of the polarization spectra of light both scattered from photosynthetic microbes and in transmission through the same cultures. For context, we also present polarization spectra of a leaf and a mineral.

There is a vast array of experimentation that may be brought to bear in the case of *in-situ* tests for the presence of biological processes (10), however *in-situ* experiments can sample only a tiny fraction of a planetary surface and its immediate subsurface, and they often anticipate a degree of specificity in the biology sought. Few remote sensing methods directly probe signatures of biological life. Trace gases can be observed that could have a biological origin, such as recent detections of localized methane production on Mars (11). Jupiter's moon Europa is strongly suspected to host a liquid water ocean (12) and infrared spectroscopic features have been shown to be consistent with those of radiation tolerant microbes (13). Beyond the Solar System, methods will be needed to assess whether extrasolar planets harbor life, and remote sensing is a necessity. Attention is being given to atmospheric composition disequilibria and to biological pigmentation spectral features as biomarkers (14-16). Typical disadvantages of these methods include model dependence and the possibility that the "biosignature" could be produced by abiotic processes, leading to a false positive.

Circular polarization may provide a more direct indication of the presence of biological processes since it is directly attributable to the chirality of the organic molecules. Earlier experiments (1, 3, 4) looked at leaves and found significant circular polarization. Here we focus on light reflected from photosynthetic bacterial cultures. Photosynthetic life must reside at the surface, use windows of atmospheric transparency and exploit regions of the spectrum where the host star shines brightly, and hence such life forms are maximally observable. The strong electronic absorption bands that are characteristic of photosynthesis are known to exhibit circular dichroism (different absorption coefficients for left- and right-circularly polarized light), hence we may anticipate a conse-quent polarization signature in scattered light, though due to the complexities of the scattering process, this is not entirely obvious in advance. For example, multiple scattering tends to randomize the polarization state; or alternatively, any circular polarization produced in the incident direction might be cancelled by circular polarization produced by subsequent reflection, as this will have the opposite handedness. If the scatterers depolarize the light, then an appropriate balance between the scattering coefficient and the differential absorption coefficient is needed to achieve measurable circular polarization. In this work, we are testing whether that balance can be achieved with microorganisms. Circular polarization can also be caused by optical interaction associated with the chirality of subcellular structures, such as membranes and macromolecules, aspects that clearly relate to the presence of biology, though yielding a distinct spectral signature relative to the circular dichroism of absorption bands (17).

Photosynthetic cyanobacteria arose between 2 and 3 billion years ago (18, 19). The enormous evolutionary advantages of photosynthesis coupled to the resilience of these microbes led to planet-wide changes in the demographic abundance of terrestrial life forms and in turn, to major changes in the composition of the Earth's atmosphere, specifically, to the rise of oxygen. That this happened early in Earth's history suggests that in an astrobiological context the occurrence and success of early microbial photosynthesis is plausibly commonplace, long-lived and statistically dominant on randomly chosen extrasolar planets. Therefore, it is possible that we may find extrasolar planets populated with organisms not unlike, though presumably not identical to, those we study here. The evolutionary advantages of photosynthesis also suggest that if life elsewhere in the Solar System has succeeded in adapting to the harsh extremes of a surface environment, then it too will likely utilize the energy consequently available through photosynthesis and phototrophy.

We therefore selected and cultured two strains of hardy chlorophyll-based photosynthetic marine cyanobacteria (20) and an α-proteobacterium that uses a different phototrophic apparatus (the bacteriochlorophyll of nonsulfur purple bacteria). A variety of antenna pigments are represented. Unicellular cyanobacteria of the genus *Synechococcus* are among the most abundant members of the picophytoplankton (20-23) and contain phycobilisome accessory light harvesting pigments. *Synechococcus* WH8101 appears green due to the presence of phycocyanin (PC) while *Synechococcus* WH7805, rich in phycoerythrin (PE), appears pink. The purple nonsulfur bacteria are of interest because of their capacity for anaerobic, non-oxygenic photosynthesis and growth with hydrogen production. We chose a versatile α-proteobacterium, *Rhodospirillum rubrum* strain ATCC 11170, possessing a well-characterized carotenoid light-harvesting apparatus, surrounding bacteriochlorophyll photocenters (24, 25).

## Results

Liquid cultures of microbes were placed horizontally in a shallow glass Petri dish, exposed to the air, approximately at room temperature. To investigate a potential connection to circular dichroism, the experiment was configured to acquire both transmission and scattered light reflection polarization spectra using a dedicated polarimeter (see materials and methods below). The sample was illuminated from above with diffuse unpolarized light for the scattering experiments. For transmission measurements, the samples were illuminated from behind by the same white light which shone onto a diffuse white plaque beneath the sample. A specialized polarimeter viewed the sample directly down the surface normal to measure the polarization of the scattered or transmitted light returned from the sample, Fig. 1.

Figs. 2–4 show the polarization spectra for our samples with one standard deviation error bars. Scaled absorption spectra are overlaid to reveal any relationships between absorption bands and circular polarization (the Cotton effect (26, 27)). The cyanobacteria, Figs. 2 & 3, show chlorophyll *a* absorption in the red ($\approx 680$ nm) and the antenna pigments phycocyanin ($\approx 620$ nm) and phycoerythrin ($\approx 560$ nm) absorption bands as expected. In Fig. 3A the blue chlorophyll *a* absorption band at $\approx 430$ nm is also visible. The polarization and absorption spectra of *R. rubrum* are shown in Fig. 3B. The primary photosynthetic reaction center molecule, bacteriochlorophyll, is seen with its absorption peak at $\approx 590$ nm, and with its 800 nm peak at the very edge of the measurement window. A mix of carotenoids causes absorption in the blue regions of the spectra. Chlorophyll *a* is absent from the *R. rubrum* spectra.

The transmission circular polarization spectra are analogous to a classical circular dichroism experiment used in protein structure and conformation analysis (28). Hence we expected to see Cotton effect circular dichroism signatures of the strong electronic absorption features and this was indeed the case. For WH8101, Fig. 2A, we see a broad absorption complex in the blue (carotenoids), and bands at 620 nm (phycocyanin) and 680 nm (chlorophyll *a*). Each shows a significant circular polarization signal. Furthermore, while the blue complex and the phycocyanin band display circular polarization of single sign, the circular polarization of the chlorophyll *a* band displays a very distinctive derivative-shaped "conservative" circular dichroism signature (29-31). The sign of the circular polarization reverses precisely at the location of the absorption maximum. This well-known effect is due to the presence of exciton-coupled chlorophyll molecular dimers where chlorophyll molecules in close proximity to one another function in pairs, effectively acting as a macromolecule.

The reflection experiment for WH8101, Fig. 2B, shows reflectance and polarization spectra that very closely mimic the transmission experiment. (The reflectance is proportional to the square of the transmittance, expected given the double pass through the material.) *The circular polarization reflection spectrum shows all the same features as the transmission polarization spectrum, even though the sample was illuminated with unpolarized light.* We see the blue-absorbing carotenoids, the phycocyanin antenna pigment and the chlorophyll *a* dimer conservative circular polarization sign-change. The amplitude of polarization is similar for transmission and reflection spectra. The presence of the circular polarization sign change at 680 nm is a very strong indication that we are witnessing the molecular circular dichroism of the material even in the scattered light spectrum.

WH7805, Fig 3A, exhibits very similar behavior, with the antenna pigment phycoerythrin absorption at $\approx 550$ nm replacing the 620 nm phycocyanin band. As before, circular polarization is present, associated with these absorption bands, including the distinctive sign change for chlorophyll. Again, the essential characteristics of the circular polarization transmission spectrum are fully reproduced in the polarization spectrum of reflected light, also shown in Fig. 3A. Likewise, Fig. 3B, the polarization reflection and transmission spectra of *R. rubrum* are similar to one another and show circular polarization associated with bacteriochlorophyll absorption at $\approx 590$ nm and blue absorbing carotenoid chromophores.

For context, Figs. 4A & B show polarization spectra of a fresh green maple leaf. The transmission spectrum shows very strong circular polarization, largely mimicking chlorophyll *a* molecular circular dichroism. The reflected circular polarization spectrum also shows a strong chlorophyll *a* 680 nm polarization signature, with an accompanying sign change through the absorption maximum. Curiously the sign is reversed with respect to the cyanobacteria, which we speculate is due to the more complex leaf optics.

Fig 4C shows the polarization of red iron oxide powder, chosen because it has a spectral edge not unlike chlorophyll and might present a false positive in a chlorophyll red-edge detection experiment. The sample also serves to guard against potential instrumental effects that might have been associated with strong spectral features (none were found). The spectropolarization signature of the iron oxide is very close to the noise limit of the instrument; there is a lack of any pronounced spectral features in circular polarization, and there is no correlation with the absorption spectrum. Likewise, Pospergelis (1) showed circular spectropolarimetry of a variety of other minerals with similar results. These characteristics are quite different to those of the biological samples.

## Discussion

Scattered light microbial polarization levels are in the range $p_c \approx 10^{-3}$ to $10^{-4}$, the leaf has $p_c \approx 2 \times 10^{-3}$, while the iron oxide has a root mean square noise level $p_c \approx 4 \times 10^{-5}$, where $p_c$ is the degree of circular polarization, which lies between 0 and 1 (see Materials and Methods below). Good astronomical polarimeters (32) can measure polarization degree $p_c \approx 10^{-6}$, though high light levels are required. Hence, the tolerance to dilution of a biological polarization signal from the many potential sources of unpolarized light that can enter the field of view of a practical remote sensing experiment is a factor $\approx 10$ to 100. For example, in arid desert environments, rocks and sand will dilute spectral features and polarization degree with unpolarized light, hence contiguous microbial colonies could be detectable if they contribute 1% to 10% or more of the light from the viewing scene. By contrast if a material is undiluted but only partially chiral, the spectral absorption band will remain at its full strength while the polarization signature is diluted due to the incomplete homochirality. (If modest chiral excesses can be found somewhere beyond the Earth, and recognized by comparison of their spectra to their polarization spectra, this would be of great interest to scientists studying the origin of life and origin of homochirality.)

Concerning the possibility of false positives, initial circular polarization imaging of the Mars surface (6) did not yield any significant signatures, which is encouraging in that false positives were not found to be abundant. Laboratory measurements of minerals, including here, have consistently revealed polarization spectra of low amplitude and different character to biological ones (1, 33). Kemp et al. (2) present circular polarization spectra of many of the Solar System bodies. A variety of effects result in polarization degree in actual abiotic situations of order $10^{-5}$ typically. The spectral dependence is smooth and slowly varying, e.g. as atmospheric scattering gives way to surface scattering on Mars from blue to red, or as in the geometric chirality of the polar effect in gas giant planets where opposite hemispheres exhibit polarization of opposite sign from scattering. The circular polarization arising from dielectric and metallic powders was investigated empirically and theoretically (33), and no significant polarization was found from the dielectric material, while the metallic powders produced circular polarization from multiple scattering. This caused variations with phase angle (viewing direction relative to incident direction) and may prove to be a useful diagnostic for metal-rich objects such as certain asteroids. The lack of pronounced spectral features and low polarization amplitude is typical of these studies, though a more extensive study of abiotic scattering coupled to empirical field polarization observations in natural environments is required to be fully confident that false positives are rare. Nevertheless these initial indications are encouraging.

On Earth, it is reasonable to expect that densely vegetated regions will produce a significant polarization signature. More challenging is the oligotrophic ocean, from which cyanobacterial sample WH7805 was derived. The oceans present a wide range of chlorophyll content, of order $0.01 - 50$ mg.m$^{-3}$. Ocean scattering and reflectance optics are dominated by chlorophyll when its concentration exceeds $\approx 0.14$ mg.m$^{-3}$ (34, 35). The chlorophyll concentration is dominated by the phytoplankton biomass, and this optical transition corresponds to a cyanobacteria density of approximately $3 \times 10^{10}$ cells.m$^{-3}$ (36). The corresponding euphotic depth is 100 m and less for higher concentrations (35). Our heuristic interpretation of the observation that the scattered and transmitted polarization levels are comparable (Figs. 2, 3) is that the polarization is produced within the layer above optical depth unity, or $\approx 0.2 \times$ the euphotic depth. Hence for oligotrophic oceanic regions whose spectral properties are dominated by the scattering and absorption from phytoplankton chlorophyll, we anticipate an implicit polarization level comparable to that of the laboratory measurements which were optimized so that the transmission optical depth $\tau \approx 1$, by dilution of a pellet containing few $\times 10^{10}$ cells. Extraneous dilution of the oceanic polarization signal will undoubtedly occur due to the presence of other particulate scatterers (though many of those will be biological), surface roughness, atmospheric scattering and clouds. Nevertheless the fact that ocean color can be dominated by chlorophyll and hence phytoplankton offers the possibility that a circular polarization degree of order $10^{-4}$ or higher may be present. Empirical measurements are clearly needed.

Elsewhere within the Solar System the ability to achieve high spatial resolution, and hence lower dilution, at high light level indicates that planetary surface polarization surveys would be feasible. For example, a detailed survey of the surface of Mars for chiral spectropolarization signatures would be possible. Circular polarization imaging of a portion of the Mars surface at two wavelengths was carried out (6) with 210 km spatial resolution using the European Southern Observatory (ESO) Very Large Telescope (VLT). While covering only a small fraction of the surface and with extremely limited wavelength coverage, their null results are encouraging as a proof of observational concept and by the absence of any false positives.

Although the prevailing opinion is that the surface of Mars is too hostile for life, Landis (37) describes the characteristics that would be required for a viable microorganism on Mars, and shows that there are terrestrial examples in the halobacteria (38). Also, very commonly terrestrial microbial life is found not highly dispersed, but in tight knit localized colonies, films and mats for protection and survival as well as to take advantage of localized niches of habitability, due to the presence of moisture and nutrients acceptable to the microbes (39, 40). Complete spectral coverage and high spatial resolution would be required to probe for localized surface microbial communities. Additionally, a potentially long racemization timescale at the surface of Mars (41) offers the possibility of seeking fossil evidence of the remnants of long-extinct biological activity using indicators of chirality. Other potential Solar System targets include Europa and Titan, as well as more primitive bodies such as comets and asteroids (42).

Eventually, with the advent of ground-based "Extremely Large Telescopes" and future dedicated space missions such as NASA's Terrestrial Planet Finder (TPF) and beyond, it will become possible to detect and characterize Earth-like extrasolar planets. The presence of biological pigments and the distinctive red-edge of chlorophyll have been proposed as potentially useful biosignatures (15, 16). Circular polarization may provide an important complement to the use of chlorophyll's red edge and in discerning whether an apparent pigmentation spectral feature has a biological origin, provided sufficient photons can be accumulated. It is unlikely that a first-generation space-based TPF mission would be able to collect enough photons for circular polarimetry, however some of the larger ground-based telescopes under consideration, such as the European Extremely Large Telescope (E-ELT) 42 m concept, would have sufficient light gathering capacity to carry out circular spectropolarimetry on Earth-like planets around some Solar neighborhood stars. (Though there will be serious issues of whether the images of these planets can be adequately separated from those of their host stars.) Beyond oxygenic photosynthesis employed by cyanobacteria and implied in the use of chlorophyll's red edge as a biomarker, it would be possible to detect enrichment of a planet by anoxygenic photosynthetic organisms. There is evidence that on the early Earth, anoxygenic photosynthetic bacteria including those that use hydrogen, hydrogen sulfide and reduced iron as electron donors preceded oxygenic photosynthesis and dominated ocean photosynthesis for more than a billion years (43), prior to the rise of oxygen in the Earth's atmosphere.

It is also plausible to consider an analogue to the circular polarization transmission measurements in which the light from the host star offers a probe through the atmosphere of a transiting planet. Hence the presence of biological molecules in the planet's atmosphere could be revealed. This may become feasible when large amplitude transits are found from planets orbiting small dwarf stars, though circular polarization from stellar magnetic fields will add to practical implementation issues for these stars.

Hence, there are situations in which the distinctive character of circular polarization spectra – the Cotton effect correlation of circular polarization with absorption bands and distinctive sign-change – could provide a very powerful indicator of the presence of optically active chiral molecules, and we conclude that under the right circumstances, circular polarization spectroscopy could be a very important tool in the search for extraterrestrial biological processes.

## Materials and Methods

### *Polarimeter*

The polarimeter (Hinds Instruments, Series II/FS42-47) is a dual photoelastic modulator (PEM) precision optical polarimeter optimized for measurement of circular polarization in the presence of significant linear polarization. If $(I, Q, U, V)$ are the usual Stokes parameters, the design goal of measuring degree of circular polarization $p_c \equiv V/I = 10^{-4}$ in the presence of linear polarization degree $p_l \equiv \sqrt{(Q^2+U^2)}/I = 0.03$ is achieved.

Light encounters two PEMs oriented 45° to one another, modulated at resonance frequencies 42 kHz and 47 kHz, followed by a Glan prism analyzer, axis 22.5° to the modulators, a field lens, monochromator and photomultiplier detector. The entire system is controlled by a dedicated desktop computer. Linear Stokes parameters, $Q$ and $U$, are measured by the *2f* modulation frequency of the PEMs, using a digital signal processor, while the circular Stokes parameter, $V$, is measured by the *1f* modulation frequency of the first PEM using a lock-in amplifier for additional precision. The DC component provides the total intensity $I$. The degree of circular polarization in the presence of fully linearly polarized light was zero within our confidence of being able to generate fully linearly polarized light. The polarimeter is tunable from 400 nm to 800 nm and is controlled automatically by software that establishes scans across a selected wavelength range in discrete steps with a specified dwell time at each wavelength. The monochromator has a spectral resolution of $\approx$ 15 nm FWHM which we most commonly sample with 5 nm step sizes.

### Microbial and Control Samples

Unicellular cyanobacteria of the genus *Synechococcus* are among the most abundant members of the picophytoplankton, which can contribute up to 30% of primary production in the surface waters of world's oceans [20, 21]. The *Synechococcus* group (Chroococcales) is a provisional assemblage loosely defined as unicellular coccoid to rod-shaped cyanobacteria, <3 μm in diameter. Many strains have been isolated from freshwater, estuarine, coastal and oceanic waters [23]. Cyanobacteria uniquely contain phycobilisome light harvesting accessory pigments, PC and PE. The PC rich *Synechococcus* dominate in coastal estuary, PE rich *Synechococcus* in the open ocean. Strains WH8101 and WH7805 represent two such picocyanobacteria [21]. WH8101 was originally isolated from the pier of Woods Hole Oceanography Institute and WH7805 from the Sargasso Sea (provided by Dr. B.J. Binder, University of Georgia). Cultures were grown in the SN medium [44] at 25°C in constant light ($\sim$25 μE m$^{-2}$ s$^{-1}$, 1 μE = 1 μmol) in an illuminated incubator. One liter of cyanobacterial culture collected at the late log phase was harvested by centrifugation (10,000 min$^{-1}$ for 15 min). Cell pellets were resuspended with 1-2 ml of SN media, and kept at 4°C until further analysis.

Purple nonsulfur bacteria are of interest for their capacity for anaerobic, nonoxygenic photosynthesis and growth with hydrogen production. We chose the versatile α-proteobacteria *R. rubrum* ATCC 11170 (Dr. G.P. Roberts, University of Wisconsin) because they possess a well-characterized carotenoid light-harvesting apparatus, surrounding bacteriochlorophyll photocenters [22]. In addition, the potential for a seminal role in the intracellular symbiosis mechanism during the formation of mitochondria by purple non-sulfur bacteria has been documented [25]. The *R. rubrum* ATCC 11170 was grown in SMN medium [45] at 25°C in 100 ml serum bottles sparged with Argon and illuminated continuously with a 100W desk lamp with rotary agitation at 2 min$^{-1}$. The *R. rubrum* was kept at its original concentration.

The fresh maple leaf is of the variety *Acer rubrum* (red maple) picked from the NIST grounds August 21, 2007. Iron oxide control (Sigma Aldrich, 12342-250G), a fine red powder that aggregates into a lumpy texture when placed in the Petri dishes (below), was used as a control.

*Measurement technique*

Each sample was placed in a shallow Pyrex glass Petri dish, with depth and density adjusted by dilution with the culture medium to yield an absorbance close to unity in the absorption bands, measured using a spectrograph (Ocean Optics). The experiments were performed in two configurations, reflection and transmission, with minimal adjustment needed to switch between the two, Fig. 1. The small amount of polarization from the light source was measured without the sample in the beam and subtracted. Each sample was scanned in wavelength typically three to five times. Multiple measurements, typically 100, were obtained at each monochromator step of the full set of Stokes parameters simultaneously. All data at a given wavelength were combined to yield the mean polarization at that wavelength. Their dispersion was used to estimate the uncertainties on the mean. For some, a small amount of smoothing was applied in the wavelength direction and the uncertainty adjusted accordingly. Error bars show one standard deviation from the resulting mean.

We derived the degree of linear polarization to check that the experiment is free of cross talk between the Stokes parameters. Scaled linear polarization curves are included in Figs. 2-4. Though linear polarization can change through absorption features it does not mimic the circular polarization. The overall wavelength dependence is much flatter, and structure associated with the chlorophyll *a* sign change seen in circular polarization is absent in the linear polarization data. We believe the changes in linear polarization are due principally to changing light paths with optical depth, as the Petri dish base is not completely flat (slightly higher in the center). The Umov effect (46) whereby linear polarization is higher in regions of low albedo may also contribute to linear polarization features.

**ACKNOWLEDGEMENT**

We acknowledge support for this work through the STScI Director's Discretionary Research fund grant #82374, and by the European Space Agency and NASA grant NNX09AC68G to S. DasSarma.

DISCLAIMER

Certain commercial equipment, instruments, or materials are identified in this paper to foster understanding. Such identification does not imply recommendation or endorsement by the National Institute of Standards and Technology nor other organizations with which the authors are affiliated, nor does it imply that the materials or equipment identified are necessarily the best available for the purpose.

REFERENCES

1. Pospergelis, M.M., (1969) Spectroscopic measurements of the four Stokes parameters for light scattered by natural objects. *Soviet Physics - Astronomy*, 12, 973-977.
2. Kemp, J.C., Wolstencroft, R.D., Swedlund, J.B. (1971) Circular polarization: Jupiter and other planets. *Nature*, 232, 165-168.
3. Wolstencroft, R. (1974) The circular polarization of light reflected from certain optically active surfaces. *Planets, Stars, and Nebulae: Studied with Photopolarimetry*, ed. Gehrels T. (University of Arizona Press, Tucson), pp 495-499.
4. Wolstencroft, R.D., Tranter, G.E., Le Pevelen, D.D. (2003) Diffuse reflectance circular dichroism for the detection of molecular chirality: an application in remote sensing of flora. *Bioastronomy 2002: Life Among the Stars*, Proceedings of IAU Symposium #213, eds Norris R., Stootman F. (Astronomical Society of the Pacific: San Francisco), p 149.
5. Barron, L.D. (2008) Chirality and life. *Space Sci Rev,* 135, 187-201. DOI 10.1007/s11214-007-9254-7.
6. Sparks, W.B., Hough, J.H., Bergeron, L. (2005) A search for chiral signatures on Mars. *Astrobiology*, 5, 737-748.
7. Cronin, J.R. and Pizzarello, S. (1997) Enantiomeric excesses in meteoritic amino acids. *Science* 275, 951–955.
8. Pizzarello, S. and Cronin, J.R. (2000) Non-racemic amino acids in the Murray and Murchison meteorites. *Geochim. Cosmochim. Acta* 64, 329–338.
9. Popa, R. (2004) *Between Necessity and Probability: Searching for the Definition and Origin of Life*. (Advances in Astrobiology and Biogeophysics, Springer).
10. Signs of Life: A Report Based on the April 2000 Workshop on Life Detection Techniques (2002); Committee on the Origins and Evolution of Life, National Research Council; The National Academies Press. ISBN 0-309-08306-0.
11. Mumma, M.J., Villaneuva, G.L., Novak, R.E., Hewagama, T., Bonev, B.P., DiSanti, M.A., Mandell, A.M., Smith, M.D. (2009) Strong Release of Methane on Mars in Northern Summer 2003. *Science*, in press. DOI: 10.1126/science.1165243
12. Kivelson, M.G., Khurana, K.K., Russell, C.T., Volwerk, M., Walker, R.J., Zimmer, C. (2000) Galileo magnetometer measurements: a stronger case for a subsurface ocean at Europa. *Science*, 289, 1340-1343.
13. Dalton, J.B., Mogul, R., Kagawa, H.K., Chan, S.L., Jamieson, C.S. (2003) Near-infrared detection of potential evidence for microscopic organisms on Europa. *Astrobiology*, 3, 505-529.
14. Woolf, N., Angel, J.R. (1998) Astronomical searches for Earth-like planets and signs of life. *ARAA,* 36, 507-537.
15. Seager, S., Turner, E.L., Schafer, J., Ford, E.B. (2005) Vegetation's red edge: a possible spectroscopic biomarker of extraterrestrial plants. *Astrobiology*, 5, 372-390.
16. Kiang, N.Y., Siefert, J., Govindjee, Blankenship, R.E. (2007) Spectral signatures of photosynthesis I review of Earth organisms. *Astrobiology*, 7, 222-251.
17. Bustamante, C., Tinoco, I., Maestre, M.F. (1983), Circular differential scattering can be an important part of the circular dichroism of macromolecules. *PNAS,* 80, 3568-3572.
18. Whitton, B. A. and M. Potts. (2000) Introduction to the cyanobacteria. In: *The Ecology of Cyanobacteria: Their Diversity in Time and Space.* Ed. B. A. Whitton and M. Potts. (Kluwer Academic Publishers, Dordrecht, The Netherlands) pp. 1-11.

19. Summons, R.E., Jahnke, L.L., Hope, J.M. (1999) 2-Methylhopanoids as biomarkers for cyanobacterial oxygenic photosynthesis, *Nature*, 400, 554-557.
20. Johnson, P., Sieburth, J. (1979) Chroococcoid cyanobacteria in the sea: a ubiquitous and diverse phototrophic biomass. *Limnol. Oceanogr.* 24, 928–35.
21. Waterbury, J. B., Watson, S.W., Guillard, R. R. L., Brand, L. E. (1979) Widespread occurrence of a unicellular, marine, planktonic cyanobacterium. *Nature* 277, 293–294.
22. Waterbury, J. B., Watson, S. W., Valois, F. W., Franks, D. G. (1986) Biological and ecological characterization of the marine unicellular cyanobacterium *Synechococcus. Can Bull Fish Aquat Sci* 214, 159–204.
23. Chen, F., Wang, K., Kan, J., Suzuki, M., Wommack E. (2006) Diverse and unique picocyanobacteria found in the Chesapeake Bay. *Appl. Environ. Microbiol.* 72, 2239-2243.
24. Papagiannakis, E., Kennis, J.T.M., van Stokkum, I.H.M., Cogdell R,J, and van Grondelle, R (2002) An alternative carotenoid-to-bacteriochlorophyll energy transfer pathway in photosynthetic light harvesting. *Proc Natl Acad Sci* 99(9), 6017–6022.
25. Cavalier-Smith, T. (2006) Origin of mitochondria by intracellular enslavement of a photosynthetic purple bacterium *Proc Biol Sci.*; 273(1596), 1943–1952.
26. Cotton A. (1895) Anomolous rotatory dispersion in absorbing bodies (in French). *Compt. Rend.* 120, 1044-1046.
27. Eliel, E.L., Wilen, S.H. (1994) *Stereochemistry of Organic Compounds* (Wiley)
28. Kelly, S.M., Price, N.C. (2000) The use of circular dichroism in the investigation of protein structure and function. *Current Protein and Peptide Science,* 1, 349-384.
29. Houssier, C., Sauer, K. (1970) Circular dichroism and magnetic circular dichroism of the chlorophyll and protochlorophyll pigments. *J Amer Chem Soc,* 92:4, 779-791.
30. Garab, G. (1996) Linear and circular dichroism. In *Biophysical Techniques in Photosynthesis, Advances in Photosynthesis,* vol. 3, eds. Amesz J., Hoff, A.J., (Kluwer) pp 11-40.
31. Blankenship, R.E. (2002) *Molecular Mechanisms of Photosynthesis* (Blackwell Science, Oxford).
32. Hough, J.H., Lucas, P.W., Bailey, J.A., Tamura, M., Hirst, E., Harrison, D., Bartholomew-Biggs, M. (2006) PlanetPol: A very high sensitivity polarimeter. *PASP,* 118, 1302-1318.
33. Degtjarev, V.S., Kolokolova, L.O. (1992) Possible application of circular polarization for remote sensing of cosmic bodies, *Earth, Moon and Planets* 57, 213-223.
34. Behrenfeld, M.J., Boss, E., Siegel, D.A., Shea, D.M. (2005) Carbon-based ocean productivity and phytoplankton physiology from space, *Global Biogeochemical Cycles*, 19, GB1006, doi:10.1029/2004GB002299.
35. Morel, A., Maritorena, S. (2001) Bio-optical properties of ocean waters: a reappraisal, *JGR*, 106, 7163-7180.
36. Glover, H.E.B., Prezelin, B.B., Campbell, L., Wyman, M. (1988) Pico- and ultra-plankton Sargasso Sea communities: variability and comparative distributions of *Synechococcus* spp. and algae. *Marine Ecology Progress Series,* 49, 127-139.
37. Landis, G.A. (2001) Martian Water: Are There Extant Halobacteria on Mars? *Astrobiology*, 1, 161-164.

38. DasSarma, S. (2006) Extreme Halophiles Are Models for Astrobiology *Microbe*, Volume 1, Number 3, p.120-127.
39. Davey, M.E., O'Toole, G.A. (2000) Microbial Biolfilms: from Ecology to Molecular Genetics, *Microbiology and Molecular Biology Reviews*, 64, 847-867.
40. Hall-Stoodley, L., Costerton, J.W., Stoodley, P. (2004) Bacterial Biofilms: from the Natural Environment to Infectious Diseases, Nature Reviews Microbiology 2, 95-108.
41. Bada, J.L., McDonald, G.D. (1995) Amino Acid Racemization on Mars: Implications for the Preservation of Biomolecules from an Extinct Martian Biota, *Icarus*, 114, 139-143.
42. Rosenbush, V., Kolokolova, L., Lazarian, A., Shakhovskoy, N., Kiselev, N. (2006) Circular Polarization in Comets: Observations of Comet C/1999 S4 (LINEAR) and Tentative Interpretation, *Icarus*, 186, 317-330.
43. Canfield, D.E., Habicht, K.S., Thamdrum, B. (1999) The Archean Sulfur Cycle and the Early History of Atmospheric Oxygen, *Science*, 288, 658-661.
44. Waterbury, J.B., Willey, J.M. (1988) Isolation and growth of marine planktonic cyanobacteria. *Methods Enzymol* 167: 100-105.
45. Kerby, R.L., Hong S.S., Ensign, S.A., Coppoc, L.J., Ludden, P.W., Roberts, G.P. (1992) Genetic and physiological characterization of the *Rhodospirillum rubrum* carbon monoxide dehydrogenase system. *J Bacteriol.* 174(16), 5284-5294.
46. Umov, N.A. (1912) Spectropolarimetric method for investigating the absorption of light and the nature of dyes (in German). *Phys. Z.*, 13, 962-971.

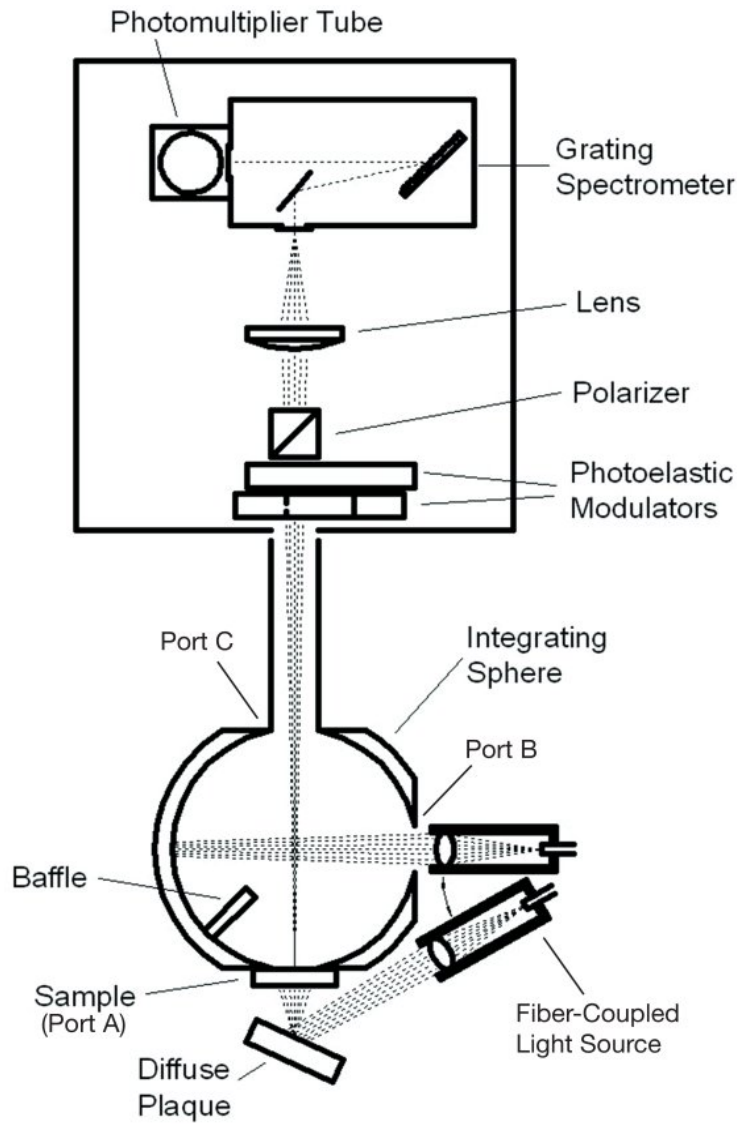

*Fig. 1. Illustration of the experimental configuration.* In reflection mode, the light, from a fiber-coupled quartz-tungsten-halogen lamp enters through an open port (B) into a 200 mm diameter integrating sphere, illuminating a spot on the sphere wall opposite. A baffle is located between the illuminated spot and the sample at port A to reduce direct illumination from that direction. The light is depolarized by numerous internal reflections and exits as unpolarized diffuse light onto the surface of the sample at port A. The polarimeter views the sample from the surface normal at port C and the sample at port A is imaged onto the entrance slit of the monochromator. In transmission mode, the horizontal port B is closed and the same light source is used to illuminate a diffuse white plaque beneath the sample and this low-polarization white light passes through the sample directly to the polarimeter.

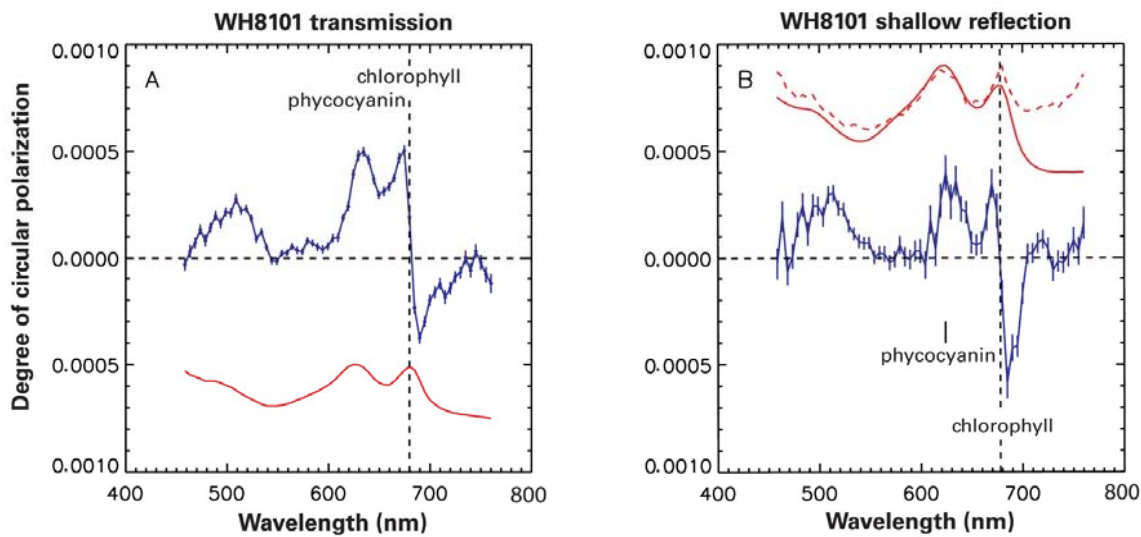

*Fig. 2. Circular polarization spectra of cyanobacteria WH8101.* Fig 2A Transmission polarization spectrum of *Synechococcus* WH8101. The blue line shows the degree of circular polarization, with ±1σ error bars; the solid red line shows a scaled version of the absorbance spectrum. Fig 2B Reflection polarization spectrum as for 2A except that the solid red line is scaled –log₁₀(Reflectance) and the dashed red line is a scaled plot of linear polarization degree.

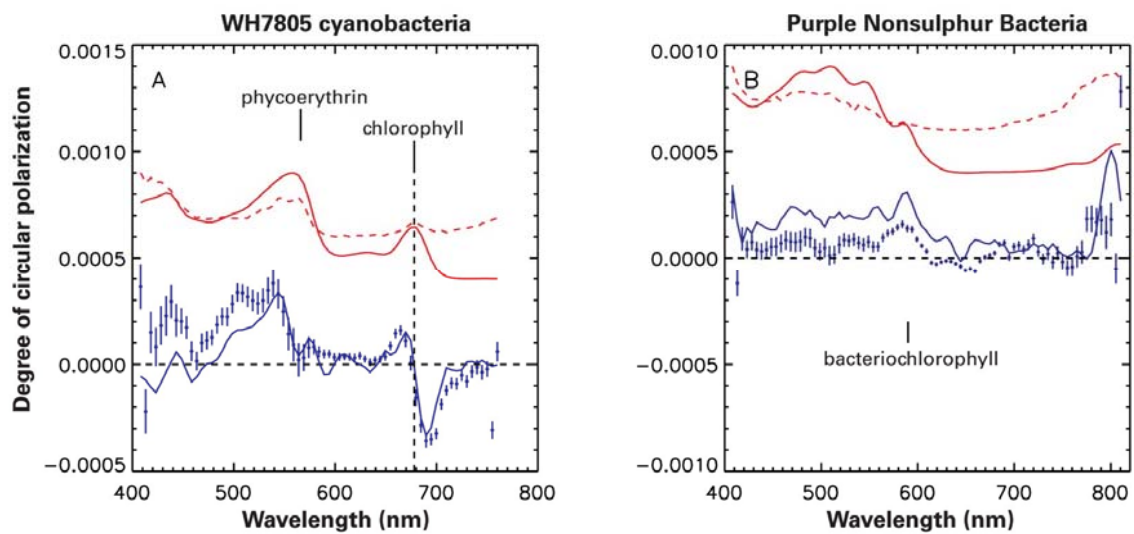

*Fig. 3. Polarization spectra of cyanobacteria WH7805 and of purple nonsulfur bacteria.* (A) Both transmission and reflection polarization spectra of cyanobacteria *Synechococcus* WH7805. Transmission is shown as a solid blue line, and reflection with solid blue circles and error bars. Note the high degree of similarity between the two. The solid red line shows a scaled -log₁₀(Reflectance) spectrum and the dashed red line a scaled plot of the degree of linear polarization. (B) Similarly shows transmission, solid blue line, and reflection, blue dots, polarization spectra for purple nonsulfur bacteria *R. rubrum*. The red solid line shows a scaled -log₁₀(Reflectance) spectrum and the red dashed line a scaled plot of the degree of linear polarization. Bacteriochlorophyll has absorption peaks at approximately 590 nm and 800 nm which appear in these data.

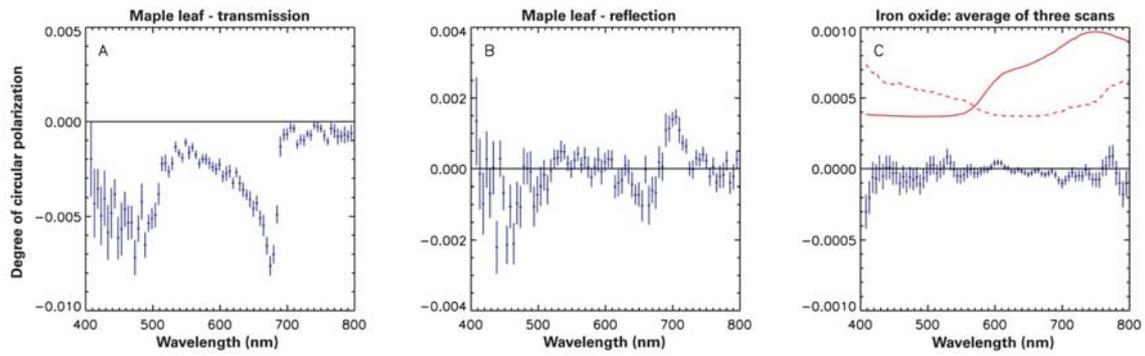

*Fig. 4. Circular polarization spectra for a leaf and a mineral.* (A) A maple leaf transmission polarization spectrum. (B) Corresponding maple leaf reflection polarization spectrum. (C) A control iron oxide polarization spectrum. The blue data points with error bars are the degree of circular polarization in each panel. The solid red line of 4C shows the reflection spectrum of iron oxide and the dashed red line, the degree of linear polarization, both arbitrarily scaled.